\documentclass[letter]{aa}
\usepackage[varg]{txfonts}

\def\sun{$_{\odot}$}

\def\12co{$^{12}$CO}
\def\co13{$^{13}$CO}
\def\NH2{N$_{\mathrm{H}_{2}}$}
\def\n2h{N$_{2}$H$^{+}$}
\def\d2n{N$_{2}$D$^{+}$}
\def\nh3{NH$_{3}$}
\def\hco{HCO$^{+}$}
\def\h13co{H$^{13}$CO}
\def\h20{H$_{2}$O}

\def\avir{$\alpha_{vir}$}
\def\0avir{$\alpha_{0}$}

\def\a_eff{$\tilde{\alpha}_{eff}$}

\usepackage{color}

\begin{document}

\title{A possible observational bias in the estimation of the virial parameter in virialized clumps}

\author{A. Traficante\inst{1}
\and Y.-N. Lee\inst{2,3}
\and P. Hennebelle\inst{3}
\and S. Molinari\inst{1}
\and J. Kauffmann\inst{4}
\and T. Pillai\inst{5,6}}

\institute{IAPS-INAF, Via Fosso del Cavaliere, 100,
              00133 Rome, Italy\\
              \email{alessio.traficante@inaf.it}
            \and
                                Institut de Physique du Globe de Paris, Sorbonne Paris Cit\'{e}, Universit\'{e} Paris Diderot, UMR 7154 CNRS, F-75005 Paris, France            
            \and 
              Laboratoire AIM, Paris-Saclay, CEA/IRFU/SAp - CNRS - Universit\'e Paris Diderot, 91191, Gif-sur-Yvette Cedex, France
            \and
             Haystack Observatory, Massachusetts Institute of Technology, 99 Millstone Road, Westford, MA 01886, USA
             \and
             Max-Planck-Institut f$\ddot{\mathrm{u}}$r Radioastronomie, Auf Dem H$\ddot{\mathrm{u}}$gel 69, Bonn, 53121, Germany 
             \and
             Institute for Astrophysical Research, Boston University, 725 Commonwealth Ave, Boston, MA 02215, USA            
            }

\abstract{The dynamics of massive clumps, the environment where massive stars originate, is still unclear. Many theories predict that these regions are in a state of near-virial equilibrium, or near energy equi-partition, while others predict that clumps are in a sub-virial state. Observationally, the majority of the massive clumps are in a sub-virial state with a clear anti-correlation between the virial parameter \avir\ and the mass of the clumps $M_{c}$, which suggests that the more massive objects are also the more gravitationally bound. Although this trend is observed at all scales, from massive clouds down to star-forming cores, theories do not predict it. In this work we show how, starting from virialized clumps, an observational bias is introduced in the specific case where the kinetic and the gravitational energies are estimated in different volumes within clumps and how it can contribute to the spurious \avir-$M_{c}$ anti-correlation in these data. As a result, the observed effective virial parameter \a_eff$<\alpha_{vir}$, and in some circumstances it might not be representative of the virial state of the observed clumps.}


\maketitle

\titlerunning{A possible observational bias in the estimation of the virial parameter}
\authorrunning{A. Traficante et al.}

\section{Introduction}
The star formation process proceeds in a hierarchical way, from clouds (size between 5-100 pc) down to cores (size $\simeq0.1-0.2$ pc), with an intermediate step, the clumps, which are cold, dense structures of $\simeq0.5-2$ pc in size. These objects have already been investigated in the past \citep[][and references therein]{Williams94}, but recently surveys of the Galactic plane such as BGPS, ATLASGAL, and Hi-GAL have identified thousands of these objects, from extremely young, starless clumps up to more evolved structures that embed HII regions \citep{Svoboda16,Csengeri14,Urquhart18,Traficante15b,Elia17,Cesaroni15}.

To understand the mechanism of star formation it is fundamental to investigate  the dynamics of the above-mentioned clumps, and a key parameter used to do that is the so-called virial parameter \avir, defined as $\alpha_{vir}=5\sigma^{2}R_{c}/(GM_{c})\propto E_{kin}/\vert E_{pot}\vert$ \citep{Bertoldi92}. Here, $E_{kin}\propto \sigma_{c}^{2}M_{c}$ and $E_{pot}\propto M_{c}^{2}/R_{c}$ are the kinetic and gravitational energy of the clump, respectively, $M_{c}$ and $R_{c}$ are the mass and radius of the clump, $G$ is the gravitational constant, and $\sigma_{c}$ is the one-dimensional (1D) velocity dispersion. Several studies of the dynamics of clumps, focusing on the more massive ones, demonstrate that \avir\ is often below the value expected for an isothermal sphere in hydrostatic equilibrium, or a Bonnor-Ebert sphere with $\alpha_{BE}\simeq2$, and in particular below the virial equilibrium value, \avir=1. In addition, each separated survey exhibits a clear anti-correlation between \avir\ and the mass of the clumps \citep{Kauffmann13,Urquhart14,Urquhart18,Traficante18}. These results have been interpreted as the more massive clumps being more gravitationally unstable and prone to collapse \citep[e.g.][]{Urquhart18}.

However, the \avir-mass anti-correlation can be also induced by some observational bias. When several surveys of different objects are combined together the trend seems to disappear, suggesting that this anti-correlation may be spurious and somehow misleading \citep{Kauffmann13}. This is corroborated by results that show that massive clumps present clear signatures of dynamical activity and infall motions, such as blue-asymmetric \hco\ spectra, regardless of the inferred value of the virial parameter or mass of the clumps \citep{Traficante17_PII,Traficante18}.

From a theoretical point of view, the turbulent-core models of massive star formation assume that massive stars form in clumps near the virial equilibrium state, supported by high levels of pressure generally driven by local turbulence \citep{McKee03,Tan14}. Recent simulations of high-mass gaseous proto-clusters have shown that these objects may undergo a local gravo-turbulent collapse and evolve at near-virial equilibrium while continuing to accrete mass \citep{Lee16a,Lee16b}, suggesting also that the energy properties of the clumps are set at the very early stages of formation. The collapse may also be driven by the gravity itself on a global scale that puts the massive clumps in a state of energy equi-partition. This implies $\alpha_{eq}=2$, a condition that in the observations, given the uncertainties of the measurements, is nearly indistinguishable from \avir=1 \citep{Ballesteros-Paredes11,Ballesteros-Paredes18,Camacho16,Iffrig17,Hennebelle18}. Alternatively, competitive accretion theories consider that massive star-forming regions may be in global, free-fall collapse, but they start and evolve in a sub-virial state, although the expected values may still be near unity \citep[\avir$\simeq0.5$,][]{Bonnell06}.

Generally speaking, the motions induced by, or nearly in equilibrium with, gravity, are naturally virialised. This is a robust property/feature from which it seems difficult to depart significantly. Even magnetic fields may not be able to provide significant support for massive clumps, from both a theoretical and observational point of view \citep[e.g.][]{Hartmann01,Crutcher12,Hennebelle18}.

All these theories can explain why clumps are in a state of near-virial equilibrium or are in a roughly sub-virial state, but none of them offer a clear explanation for the observed anti-correlation between \avir\ and mass, reinforcing the idea that it may be the consequence of an observational bias in the evaluation of the gravitational energy or the kinetic energy (or both) at all spatial scales. This \avir-mass anti-correlation is in fact observed with similar trends in giant clouds \citep[GMCs,][]{Miville-Deschenes17}, clumps \citep{Traficante18} and cores \citep{Kauffmann13}. However, the bias may have different origins. The physical and kinetic properties of giant clouds are usually determined consistently within the same volume of gas from the emission line spectra of CO isotopologues \cite[e.g.][]{Heyer09,Roman-Duval10,Miville-Deschenes17}. On parsec scales, within the dense clumps \citep[$\Sigma\geq0.01$ g cm$^{-2}$][]{Elia17,Urquhart18}, the CO lines become optically thick and the physical properties must be determined from the cold dust emission, and then combined with the kinematics derived from the gas to evaluate the virial properties.

A possible bias in the estimation of the gravitational energy in massive regions is discussed in the literature, and it is the sensitivity limit of each survey that may bias the observations towards regions with similar column densities \citep{Ballesteros-Paredes12,Kauffmann13}, which is particularly significant for relatively low-density regions like the GMCs.

In this letter we show that, starting from a simple model of massive clumps near virial equilibrium, an observational bias is introduced when the gravitational and the kinetic energies are evaluated within different regions of the clump and that this bias can contribute to the spurious anti-correlation between \avir\ and mass. The model is described in Sect. \ref{sec:model}, and its application to simulations of cluster formation and real datasets is reported in Sects. \ref{sec:simulations} and \ref{sec:data_results}, respectively. In Sect. \ref{sec:conclusions} we draw our summary. More details about the setup of the simulations are in Appendix \ref{app:simulations}, and a discussion of the possible biases introduced in observations of massive clumps is given in Appendix \ref{app:discussion}.

\section{A bias in the observations of virialized clumps}\label{sec:model}
We model a clump as an `onion-like' structure, that is, a spherically symmetric region formed by the combination of several layers of material with increasing density as we move from the clump edges into the inner regions.

Each clump therefore follows a density profile $\rho$ that depends only on the radius $R$ as $\rho=\rho_{0}(R/R_{c})^{-p}$, with $\rho_{0}$ the density at the edges of the clump, that is, at the layer with radius $R=R_{c}$. 

We further assume that the clumps are all near virial equilibrium at all stages of star formation and that each layer is virialized itself, that is, clumps have \avir$\simeq1$ at each radius $R$. We note that our model is not intended to reconstruct the absolute value of the virial parameter in massive clumps. For the purpose of our discussion, it is not relevant if the clumps are all in virial equilibrium (\avir=1), can be modeled as a Bonnor-Ebert sphere ($\alpha_{BE}\simeq2$), or are in energy equi-partition ($\alpha_{eq}=2$), as long as they all have the same intrinsic value of $\alpha$.

Finally, we assume that the clumps follow the relation  \citep{Camacho16,Ballesteros-Paredes18,Iffrig17}

\begin{equation}\label{eq:Larson_generalized}
\sigma\propto (\Sigma R)^{0.5},
\end{equation}

\noindent with $\Sigma$ the surface density of the clumps. This is a generalized version of the first Larson's relation \citep[$\sigma\propto R^{0.5}$][]{Larson81,Heyer04} that accounts for the results obtained in recent observations of massive objects \citep{Ballesteros-Paredes11,Maud15,Traficante17_PII} and allows us to self-consistently assume that clumps are in virial equilibrium \citep{Ballesteros-Paredes18}. Equation \ref{eq:Larson_generalized} implies $\sigma\propto R^{\delta}$ with $\delta>0$ when $       \Sigma=M/(\pi R^{2})\propto R^{\beta}$ and $\beta>-1$, that is, a mass-radius relationship $M\propto R^{\gamma}$ with $\gamma>1$, a condition that is always found in observations \citep[e.g.][]{Urquhart14,Ellsworth-Bowers15,Traficante17_PII}.

This toy-model is an approximation of the real structure of the clump, where low-density gas can be widely distributed across all spatial scales (see Sect. \ref{sec:simulations}). Also, it may be inaccurate to describe the central regions of the clump, where the deep potential well may allow the fragmentation and the consequent formation of many non-uniformly distributed cores and protostars \citep[e.g.][]{Zhang15}. Nevertheless, it is a very simple but reasonable approximation for clumps that have an aspect ratio of the order 1.2-1.3 \citep{Molinari16_cat}, and in particular to model the outer layers of each clump, which is the primary purpose of this work.

Applying the virial equilibrium hypothesis to a clump,  and following the definition of the virial parameter of \citet{Bertoldi92}, we can write 

\begin{equation}
\frac{GM_{c}}{5R_{c}}=\sigma_{c}^{2}.
\end{equation}

The quantities $M_{c}$ and $R_{c}$ in the clumps are determined by observations of the dust continuum emission in the far-infrared/sub-millimetre \citep{Urquhart14,Urquhart18,Traficante15b,Svoboda16,Elia17}. In order to evaluate the kinetic energy of the clumps, instead observers usually measure the velocity dispersion from the emission spectra of a specifically optically thin line such as for example \nh3 $(1,1)$ \citep{Urquhart14} or \n2h\ $(1-0)$ \citep{Jackson13}. Each particular transition however is excited at and above a specific volume density, the critical density of the line $\rho_{crit}$. Although also including the effect of the radiative trapping, \citet{Shirley15} showed that the effective excitation density $\rho_{eff}$ may be significantly lower than the line critical density of a given gas tracer, the gas at densities lower than $\rho_{eff}$ is not traced by the chosen molecular line \citep[see e.g.][]{Kauffmann17}. In this case, there will be a specific radius $R_{eff}$ above which the gas is not traced by our chosen molecule. The measured velocity dispersion, $\sigma_{eff}$, refers therefore to a region of the clump that goes from the center up to the layer of the clump where $M=M_{eff}$ and $R=R_{eff}<R_{c}$. Since in our hypothesis the clumps are in virial equilibrium at all layers, at a given radius $R_{eff}<R_{c}$ we can write

\begin{equation}\label{eq:m_eff_sigma_eff}
\frac{GM_{eff}}{5R_{eff}}=\sigma_{eff}^{2} < \sigma_{c}^{2},
\end{equation}

\noindent with $\sigma_{eff}^{2} < \sigma_{c}^{2}$ , which comes from our hypothesis that the clumps follow a generalized Larson-like relation $\sigma\propto R^{\delta}$ with $\delta>0$ and $R_{eff}<R_{c}$.

The key point is that for each clump we can only determine the total mass $M_{c}$ and radius $R_{c}$ but, at the same time, for any given molecule we can only measure the velocity dispersion $\sigma_{eff}$. Although the clumps are all in virial equilibrium, the net result is that we measure an \textit{effective virial parameter} \a_eff: 

\begin{equation}\label{eq:alpha_eff_funct}
\tilde{\alpha}_{eff}=\frac{5\sigma_{eff}^{2}R_{c}}{GM_{c}}<\frac{5\sigma_{c}^{2}R_{c}}{GM_{c}}=\alpha_{vir}=1.
\end{equation}

The value of \a_eff depends on the ratio between the kinetic energy measured at and within the region with effective critical density $\rho_{eff}$, proportional to $\sigma_{eff}^{2}$, and the total kinetic energy of the clump (proportional to $\sigma_{c}^{2}$). 

In the following, we derive the dependence of \a_eff\ with the effective critical density $\rho_{eff}$ within each clump (Sect. \ref{sec:a_eff_rho_eff}) and the slope of the \a_eff-mass diagram for an ensemble of clumps (Sect. \ref{sec:a_eff_mass_sample}).

\subsection{The relation between \a_eff\ and $\rho_{eff}$ in a given clump}\label{sec:a_eff_rho_eff}

In our hypothesis each layer is virialized, and following on from Eqs. \ref{eq:m_eff_sigma_eff} and \ref{eq:alpha_eff_funct}:

\begin{equation}\label{eq:alpha_eff_mass_radius}
\frac{\tilde{\alpha}_{eff}}{\alpha_{vir}=1}=\frac{\sigma_{eff}^{2}}{\sigma_{c}^{2}}=\frac{M_{eff}}{R_{eff}}\frac{R_{c}}{M_{c}}.
\end{equation}

As showed in Appendix \ref{app:equation}, given the definition of the effective critical density $\rho_{eff}=\rho_{0}(R_{eff}/R_{c})^{-p}$, we can express the effective virial parameter measured in a given clump as a function of $\rho_{eff}$:

\begin{equation}\label{eq:alpha_eff_density_relation}
\tilde{\alpha}_{eff}(\rho_{eff})=\bigg(\frac{3}{3-p}\frac{\rho_{eff}}{\overline{\rho}_{c}}\bigg)^{\frac{p-2}{p}},
\end{equation}

\noindent where $\overline{\rho}_{c}$ is the average volume density of the clump. This equation shows that for a clump with a given density profile, when $p\neq2$ there will only be some specific molecule with a critical density $\rho_{eff}=\rho_{0}=\overline{\rho}_{c}\times(3-p)/3$ for which $\tilde{\alpha}_{eff}=\alpha=1$. With $p\neq2$ and in clumps with $\overline{\rho}_{c}<\rho_{eff}$, $\tilde{\alpha}_{eff}<\alpha_{vir}$ and it depends on the chosen gas tracer used to infer the kinetic properties of the gas.

Similarly, Eq. \ref{eq:alpha_eff_density_relation} can be expressed as a function of the observational quantity $\sigma_{eff}$. From Eq. \ref{eq:alpha_eff_mass_radius} it follows that

\begin{equation}\label{eq:sigma_eff_density_relation}
\sigma_{eff}(\rho_{eff})=\sigma_{c}\bigg(\frac{3}{3-p}\frac{\rho_{eff}}{\overline{\rho}_{c}}\bigg)^{\frac{p-2}{2p}}
.\end{equation}

The radial density profile $p$ has a range of values both from theoretical and observational points of view. If the clump is described by a logatropic Equation of state, it follows that $p=1$ \citep{McLaughlin97}. On the other hand, the "inside-out" collapse model of a singular isothermal sphere predicts $p=2$ \citep{Shu77}. The core-collapse model for the formation of massive stars proposed by \citet{McKee03} assumes an intermediate value $p=1.5$. In massive star-forming clumps, $p$ has been observed in the range $p=1.6\pm0.5$ \citep{Beuther02a}, or $p=1.8\pm0.4$ \citep{Mueller02}. The density profile of massive clumps derived by their surface density profile suggests $p=1.1$ \citep{Tan13}. 

With a value of $p=1.5$, Equation \ref{eq:sigma_eff_density_relation} predicts $\sigma_{eff}\propto\rho_{eff}^{-0.17}$, in reasonable agreement with observations in nearby high-mass star-forming regions \citep[$\sigma_{eff}\propto \rho^{-0.15}$,][]{Orkisz17}.

\subsection{The \a_eff-mass relation for an ensemble of clumps}\label{sec:a_eff_mass_sample}
In order to derive the \a_eff-mass relation that resembles the observations, Eq. \ref{eq:alpha_eff_density_relation} has to be applied to a sample of clumps observed with a single gas tracer. In this case, $\rho_{eff}$ is defined for the entire sample and the variable becomes $\overline{\rho}_{c}$. For a given gas tracer, there will only be a class of clumps with average volume density $\overline{\rho}_{c}=3/(3-p)\times\rho_{eff}$ for which we measure \a_eff=1. The value of $\overline{\rho}_{c}=3M_{c}/(4\pi R_{c}^{3})$ of each clump depends on its mass $M_{c}$ and radius $R_{c}$; therefore the distribution of $\overline{\rho}_{c}$ is directly linked to the mass-radius distribution of the sample, which can be expressed as $M_{c}\propto M_{0}(R_{c}/R_{0})^\gamma$. The factors $M_{0}$ and $R_{0}$ are normalization values that may depend on the various techniques used to derive the dust properties in each survey of clumps. It follows that

\begin{equation}
\overline{\rho}_{c}(M_{c})=\frac{3}{4\pi}\frac{M_{c}}{R_{c}^{3}}=\frac{3}{4\pi}M_{0}^{\frac{3}{\gamma}}R_{0}^{-3}M_{c}^{1-\frac{3}{\gamma}}.
\end{equation}

Substituting this value into Eq. \ref{eq:alpha_eff_density_relation}, the effective virial parameter measured in a sample of clumps can finally be expressed as function of the mass of the clumps:

\begin{equation}\label{eq:alpha_vir_mass_clump_relation}
\tilde{\alpha}_{eff}(M_{c})=A M_{c}^{\frac{\gamma-3}{\gamma}\frac{2-p}{p}}=A M_{c}^{h},
\end{equation}

\noindent with 

\begin{equation}\label{eq:alpha_vir_mass_clump_relation_coeff}
A=\bigg[\rho_{eff}\frac{4\pi}{3-p}\bigg(\frac{R_{0}^{\gamma}}{M_{0}}\bigg)^{\frac{3}{\gamma}}\bigg]^{\frac{p-2}{p}}.
\end{equation}

The bias induced by the use of a single gas tracer to measure the kinetic energy of a sample of virialized clumps induces a \textit{spurious} correlation between \a_eff\ and $M$ that depends on two parameters: the slope $\gamma$ of the mass-radius relationship of the sample and the average radial density profile $p$ of the clumps. In particular, with a mass-radius slope of $\gamma<3$ and a clump density profile $p<2$, Eq. \ref{eq:alpha_vir_mass_clump_relation} predicts that we will introduce an anti-correlation between the measured virial parameter and the mass of the sample. We note that the slopes $\gamma$ and $p$ are not necessarily dependent variables. This would only be true if clumps of different sizes were representative of the structures of a single clump at different radii, which is not necessarily the case.

The value of the mass-radius slope in star-forming clumps is expected to be $\gamma\leq2$ and depends on the slope of the column density PDF of the sample \citep{Ballesteros-Paredes12}. The mass-radius diagram has been evaluated for several surveys of clumps and the derived $\gamma$ depends significantly on the approach used to extract the parameters of clumps and to fit the mass-radius slope \citep{Kauffmann10b}. Also, different algorithms may lead to different values of the column density PDF and of the mass-radius slope \citep[][]{Gomez14}. Mass-radius diagrams of nearby star-forming regions and massive clumps across the Galaxy have values in the range $1.4\leq\gamma\leq1.7$  \citep{Kauffmann10b,Urquhart14,Urquhart18}. There are however examples in the literature of samples of clumps that exhibit $\gamma\gtrsim2$ \citep{Ellsworth-Bowers15}, larger than 2 \citep[$\gamma=2.38$,][]{Traficante18} or even steeper \citep[$\gamma\simeq2.7$,][]{Kainulainen11}.

\begin{figure}
\centering
\includegraphics[width=8cm]{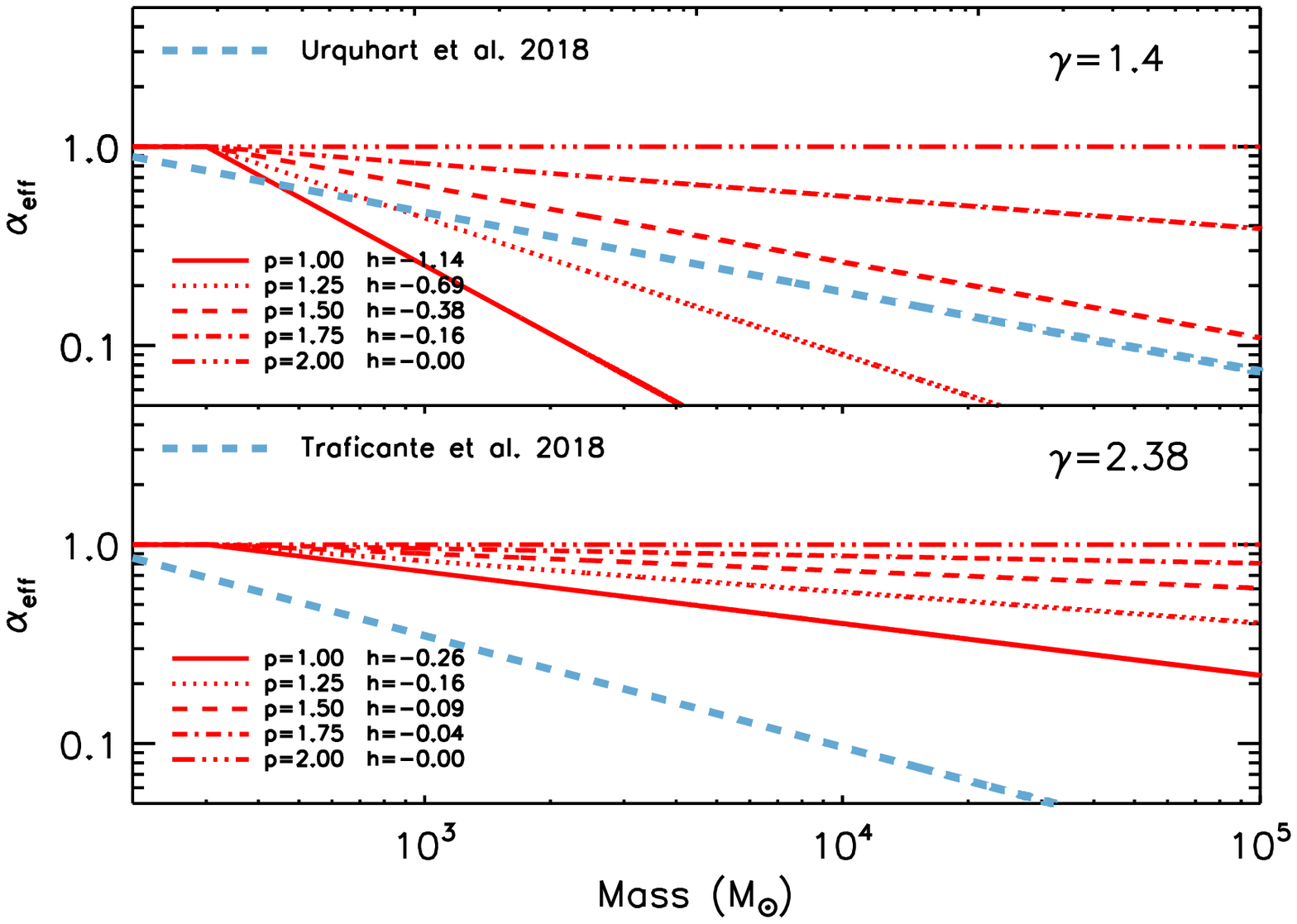}
\caption{Prediction of the model described by Eq. \ref{eq:alpha_vir_mass_clump_relation} for different values of $p$ (various red lines) and for a mass-radius slope $\gamma=1.4$ (upper panel) and $\gamma=2.38$ (lower panel), using results obtained in the samples of \citet{Urquhart18} and \citet{Traficante18}, respectively. The models have been normalized to all have \a_eff=1 at $M_{0}\leq300$ M\sun. The azure-dashed lines are the \a_eff-mass relations found in \citet[][upper panel]{Urquhart18} and \citet[][lower panel]{Traficante18} respectively, normalized to have \a_eff=1 at $M_{0}=150$ M\sun\ for ease of visualization.}
\label{fig:avir_mass_model_clumps}
\end{figure}

In Fig. \ref{fig:avir_mass_model_clumps} we show the predicted value of the slope $h$ for different values of $p$ in the expected range $1<p<2$ and for two values of the mass-radius slope, $\gamma=1.4$ \citep[][upper panel]{Urquhart18} and  $\gamma=2.38$ \citep[][lower panel]{Traficante18}. The models have been normalized to have \a_eff=\avir=1 in clumps with mass below $M_{0}=300$ M\sun\ for all values of $p$. Assuming a clump with $M=M_{0}$, a radius $R_{0}=0.4$ pc,  and a density profile with $p=1.5$, it follows from Eqs. \ref{eq:alpha_vir_mass_clump_relation} and \ref{eq:alpha_vir_mass_clump_relation_coeff} that \a_eff=\avir=1 is obtained if the kinetic energy is measured using a gas tracer with effective critical density $\rho_{eff}=10^{4}$ cm$^{-3}$, like for example the \n2h\ $(1-0)$ \citep{Shirley15}. The figure also shows the \a_eff-mass slopes measured in the work of \citep{Urquhart18} and \citep{Traficante18}, in the upper and lower panels, respectively, normalized to \a_eff=\avir=1 at $M_{0}=150$ M\sun\ for ease of visualization. We note that the absolute value of \a_eff\ depends on the particular survey, but within each survey the measurements are consistent and the \a_eff-mass slope can be compared with the models.

In the following sections we see how this simple model explains in part what is found in simulations and in observations of massive regions.

\section{The simulations}\label{sec:simulations}
The previous section provided a simple model to explain the bias in velocity dispersion that could be introduced by molecular line observations. 
In this  section, we test this idea in simulated cluster formation to confirm the model.

The simulations from \citet{Lee16a} and with more complete physics, including ionizing feedback, are used to test and analyse the concepts outlined above. They all show qualitatively similar results. Therefore, here we present results from run B in \citet{Lee16a}, which is a single cluster of a few thousand solar masses formed from the collapse of a $10^{4}$ M\sun\ molecular cloud. The cluster has a radius of 0.6 pc. These simulations are useful to illustrate the effect of the observational bias in a single protocluster. They represent only one point in the \avir-mass space and were not designed to reproduce the \avir-mass relation. More details about the simulations used in this work can be found in Appendix \ref{app:simulations}.

In Fig. \ref{fig:sim_alpha_vir_radius} we show the effective virial parameter computed at each radius and for several values of $\rho_{eff}$. This value is simply defined as $\alpha_{eff} = 5/3\sigma_{3D}^{2}(R,\rho>\rho_{eff}) \times R / (GM(R))$, where $\sigma_{3D}(R,\rho>\rho_{eff})$ is the 3D velocity dispersion of the gas within radius $R$ and with density $> \rho_{eff}$. The factor $5/3$ makes the results obtained from the simulations consistent with the definition of \avir\ given by \citet{Bertoldi92}, and directly comparable with the observations, which considers the 1D velocity dispersion. When most of the mass is considered, the cluster is close to viral equilibrium, that is, \a_eff$\sim $ \avir$\sim 1$. We note that there is a numerical constant from geometrical considerations that we ignore here since it does not have a qualitative impact on the results. 

When the low-density gas is filtered out, that is, at increasing values of $\rho_{eff}$, we see a significant decrease in $\alpha_{eff}$ in regions up to twice the cluster size (radius of $\simeq1.2$ pc), which is in line with the results by \citet{Orkisz17} and the \avir-mass trend reported by several cluster observations. At larger radii the dense gas becomes a very small fraction of the total gas, the simulations deviate significantly from our proposed toy model and the kinematics determined with high-density tracers are not well defined. We caution that the simulations were not meant to reproduce the simple `onion' model we propose in this work, but they actually result in a more clumpy cluster than the onion shell model, and the dense gas is more widely distributed. This could possibly be regarded as several local `onions' packed together. More careful studies are certainly needed for a better understanding of the kinetics of the cluster, but these results already suggest that the measured \a_eff\ in dense clumps may be biased by the chosen gas tracer.

\begin{figure}
\centering
\includegraphics[width=8cm]{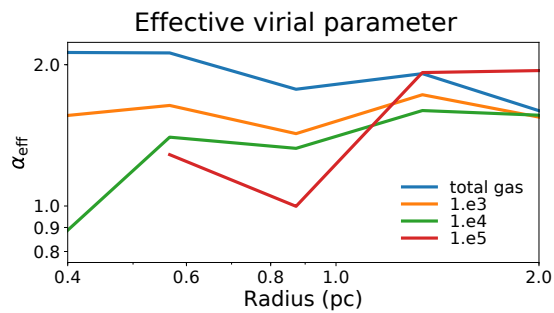} 
\caption{Effective virial parameter as a function of the radius of the cluster centre, computed for several values of the effective critical density of a given tracer: $\rho_{eff}=(10^{3}, 10^{4}, 10^{5})$ cm$^{-3}$ for the orange, green, and red lines, respectively. The blue line is the total mass of the gas.} 
\label{fig:sim_alpha_vir_radius}
\end{figure}

\section{Comparison with observations}\label{sec:data_results}

The observed \a_eff-mass anti-correlation in clumps has the form \a_eff$\propto M^{h_{\alpha}}$, with slope $h_{\alpha}$. We note that $h=h_{\alpha}$ only in the case in which the bias discussed in this work is the only factor responsible for the observed \a_eff-mass slope. We further discuss this  point in Appendix \ref{app:discussion}.

In Fig. \ref{fig:avir_gamma} we show the expected value of the slope $h$ as obtained from Eq. \ref{eq:alpha_vir_mass_clump_relation}, for a range of mass-radius slopes ($1\leq\gamma\leq3$) and for three different values of the radial density profile, $p=[1.0, 1.5, 2.0]$ (red, blue, and green lines, respectively). 

The filled circle and diamond points are the  $h$  values of the clumps taken from \citet[][and adapted from \citet{Wienen12} and \citet{Sridharan05}, respectively]{Kauffmann13}. These clumps have masses in the range $50\leq M\leq10^{4}$ M\sun, and the kinetic energy is estimated from the \nh3\ $(1,1)$ line width \citep{Wienen12,Sridharan05,Kauffmann13}. The derived \a_eff-mass slopes are $h_{\alpha}=-0.43$ and $h_{\alpha}=-0.47,$ respectively. The filled square is the $h$ value obtained from the distance-limited ensemble of $\simeq5000$ ATLASGAL clumps where the kinetic energy is estimated combining several surveys of gas tracers including \nh3 $(1,1)$, C$^{18}$O $(1-0),$ and \n2h\ $(1-0)$ \citep[][and references therein]{Urquhart18}. If the bias in the estimation of \a_eff\ is entirely due to the measurement of the kinetic energy, implying $h=h_{\alpha}$, the density profiles of the clumps in these surveys predicted by our model would be in the range $1.2\leq p\leq1.5$, in good agreement with observations  \citep{Beuther02a,Mueller02,Tan14}. The hourglass and triangle points are the values derived from the survey of \citet{Traficante18} and a collection of nearby massive regions discussed in \citet{Kainulainen11}, respectively. These surveys span a range of masses $10^{-1}\leq M\leq10^{4}$ M\sun\ and both have a relatively high mass-radius slope, with values of $\gamma=2.38$ and $\gamma=2.7,$ respectively. If $h=h_{\alpha}$, this would imply a radial density profile of the clumps with $p=0.69$ and $p=0.33,$ respectively, slopes that are below the expected values of the profile of the clumps. With a more realistic density profile in the range $1\leq p\leq1.5$, from Eq. \ref{eq:alpha_vir_mass_clump_relation} we still obtain an anti-correlation between \a_eff\ and $M$ with slope $h$, but in this case $\vert h\vert<\vert h_{\alpha}\vert$. The observed \avir-mass anti-correlation may in fact be the result of several sources of bias, as we discuss in Appendix \ref{app:discussion}.

\begin{figure}
\centering
\includegraphics[width=8cm]{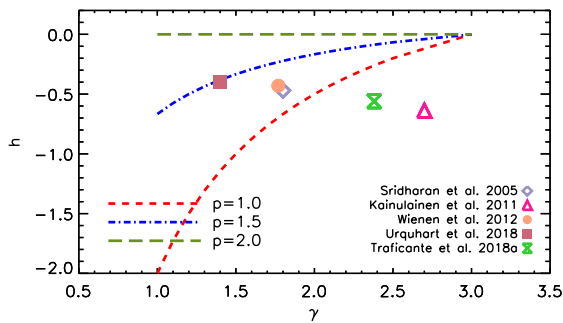} 
\caption{Predicted values of the \a_eff-mass slope $h$ for a range of values of the mass-radius slope $\gamma$ and for three different values of the radial density profile $p$ following Eq. \ref{eq:alpha_vir_mass_clump_relation}. Red, blue, and green dashed lines are the curves for values of p=[1.0,1.5,2.0], respectively. The values of $\gamma$ and $h$ from several surveys of clumps presented in the literature have been superimposed.}
\label{fig:avir_gamma}
\end{figure}

\section{Summary}\label{sec:conclusions}
In this letter we showed that, for a sample of virialized clumps modelled as an `onion-like' sphere made of several virialized layers of material, the derivation of the physical and kinematic properties from different tracers (e.g. from dust emission and molecular lines, respectively) leads to a biased estimate of an effective virial parameter \a_eff<\avir=1 and a spurious anti-correlation between \a_eff\ and the mass of the clumps. We compared our model with predictions from the simulations of \citet{Lee16a} and we showed that, although the onion-like model is a simplification of the structure of a clump, the value of the effective virial parameter depends on the different density thresholds probed by each gas tracer. 

We stress that an anti-correlation between mass and the virial parameter is observed at all scales and also in cases where the kinetic and the gravitational energies are observed within the same volumes, as for example in GMCs. The model we propose here applies under the specific conditions typical of the observations of massive clumps.

We conclude that the bias discussed in this letter can contribute at least partly, and in some cases significantly, to the \a_eff-mass slope $h_{\alpha}$ observed in several surveys of massive clumps. At the same time, under the simple onion-like approximation for the clump structure, Eq. \ref{eq:alpha_eff_density_relation} provides a simple tool to identify the best tracer required to observe the gas kinematics of a given region with average volume density $\overline{\rho}_{c}$ and a defined density profile.

Finally, it is worth noting that the conclusions of this work remain consistent for any fixed value of \avir\ in a sample of clumps and the analysis we have shown here does not aim to distinguish among different star-formation theories.

\begin{acknowledgements}
The authors want to thank the anonymous referee that helped to improve the quality of the manuscript with his/her suggestions; this work has benefited from research funding from the European Community's Seventh Framework Programme. 
\end{acknowledgements}

\bibliographystyle{aa} 
\bibliography{Final_arxiv.bbl} 

\appendix

\section{Derivation of the relation between $\tilde{\alpha}_{eff}$ and $\rho_{eff}$}\label{app:equation}
In this section we demonstrate how to derive the relation between $\tilde{\alpha}_{eff}$ and $\rho_{eff}$ starting from Eq. \ref{eq:alpha_eff_mass_radius}.

First, we recall that, given our onion-like model, the mass $M_{eff}$ at a given radius $R_{eff}$ can be expressed as 

\begin{equation}\label{eq:mass_from_density}
M_{eff}=4\pi \rho_{0}R_{c}^{p}\int_{0}^{R_{eff}}R^{2-p}dR.
\end{equation}

Solving the integral we obtain

\begin{equation}\label{eq:mass_clump_from_density_solution}
M_{eff}=\frac{4\pi\rho_{0}R_{c}^{p}}{3-p}R_{eff}^{3-p},
\end{equation}

\noindent and substituting in Eq. \ref{eq:alpha_eff_mass_radius},

\begin{equation}\label{eq:alpha_eff_radius_mass}
\tilde{\alpha}_{eff}=\frac{4\pi}{3-p}\rho_{0}R_{eff}^{2-p}\frac{R_{c}^{p+1}}{M_{c}}.
\end{equation}

The effective virial parameter can now be expressed as a function of the effective critical density $\rho_{eff}=\rho_{0}(R_{eff}/R_{c})^{-p}$. Equation \ref{eq:alpha_eff_radius_mass} becomes

\begin{equation}\label{eq:almost_alpha_eff_rho_eff}
\tilde{\alpha}_{eff}=\frac{4\pi}{3-p}\rho_{0}^{\frac{2}{p}} \rho_{eff}^{\frac{p-2}{p}}\frac{R_{c}^{3}}{M_{c}}=\frac{3}{3-p}\rho_{0}^{\frac{2}{p}} \rho_{eff}^{\frac{p-2}{p}}\overline{\rho}_{c}^{-1},
\end{equation}

\noindent with $\overline{\rho}_{c}=3M_{c}/(4\pi R_{c}^{3})$ being the average volume density of the clump. 

The volume density at the outer layer of the clump, $\rho_{0}$, can be derived from Eq. \ref{eq:mass_clump_from_density_solution} substituting $R_{eff}$ with $R_{c}$, and $M_{eff}$ with $M_{c}$. It follows that

\begin{equation}\label{eq:rho_0_rho_average_relation}
\rho_{0}=\overline{\rho}_{c}\frac{3-p}{3},
\end{equation}

\noindent and substituting in Eq. \ref{eq:almost_alpha_eff_rho_eff} we finally obtain the relation between $\tilde{\alpha}_{eff}$ and $\rho_{eff}$ described in Eq. \ref{eq:alpha_eff_density_relation}.

\section{Simulations}\label{app:simulations}
The simulations of molecular cloud collapse used in this work are initialized with a Bonner-Ebert sphere and seeded turbulence following the Kolmogorov spectrum with random phases, which initially virializes the cloud. 
The cloud collapses under its self-gravity and a gaseous proto-cluster of sub-parsec size is formed at the centre. This proto-cluster is in virial equilibrium, with the turbulence being the major support against self-gravity. Analyses by \citet{Lee16a} showed that the thermal and magnetic energies are at percent level with respect to the turbulence at this scale. 
This is a natural consequence seen in many collapsing structures: concentrated mass supported by accretion-driven turbulence \citep[see e.g.][]{Klessen10,Lee16b}. 
The first interpretations of these simulations already strongly suggest that it is very unlikely to see sub-virial structures when self-gravity is in action. Several setups were considered, with the mass of $10^3$, $10^4$, and $10^5$ solar masses. The initial density was also varied, which corresponded roughly to the scatter around Larson's relations. 

The results from the simulations are taken at a time when the collapsing region of the proto-clusters is not overly dominated by the sink-particles, that is, when the mass of the sink particle is less than half of the initial proto-cluster mass. We note that similar results are obtained if the simulations are taken at earlier stages, while at later stages the gas motions are overly affected by the feedback of the already formed HII regions.

The radial-density profiles in these simulations is roughly $\rho\propto r^{-2}$, however in the simulations the substructures are present across the clump and not only in the central region, as instead suggested by our toy model. As discussed in the main text of this letter, these simulations were not designed to reproduce our results, but to get a first glance of the model described in this work.

\begin{figure}
\centering
\includegraphics[width=8cm]{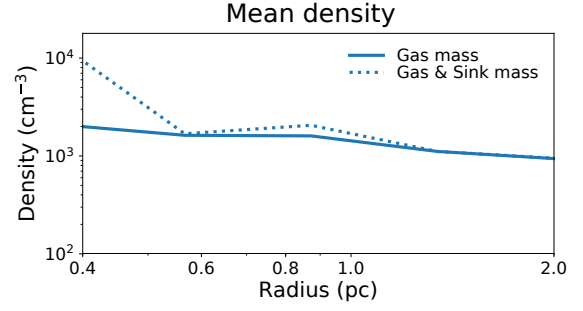} 
\caption{Mean density of the cluster center simulated in \citet{Lee16a} averaged in spherical shells against the radius of the cluster. The Figure shows how the density increases towards the center of the cluster with (dashed line) and without (solid line) including the mass of the sink particles.}
\label{fig:sim_density_radius}
\end{figure}

\begin{figure}
\centering
\includegraphics[width=8cm]{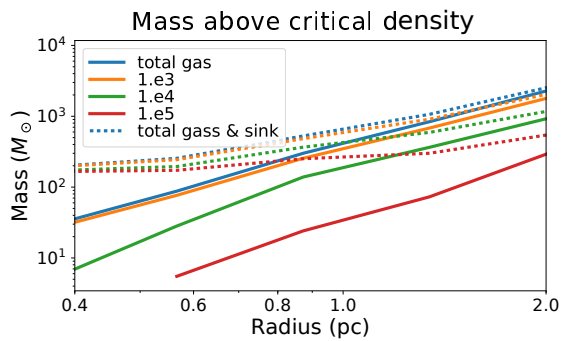} 
\caption{Mass integrated above a fixed value of the effective critical density $\rho_{eff}=(10^{3}, 10^{4}, 10^{5})$ g cm$^{-3}$ for the orange, green, and red lines, respectively, as function of the radius of the cluster centre. The blue line is the total mass of the gas. Dashed lines include also the mass of the sink particles.}
\label{fig:sim_mass_density}
\end{figure}

Figure \ref{fig:sim_density_radius} shows the density averaged in spherical shells against the distance to the cluster centre,  with the solid line representing the gas mass and the dotted line representing the total mass including sink particles for completeness. The mass $M$ evaluated in these simulations does not include the mass of the sink particles. This is done to extract  parameters from
the simulations that are as close as possible to the observed values, since sink particles are dominated by the stars that are not observed in the dust continuum surveys used to determine the clump physical properties. The resolution of these simulations is however $\simeq200$ AU, the size of a region that still contains some gas surrounding the final stars. Higher-resolution simulations able to resolve single stars starting from parsec-scale molecular clouds are significantly more time consuming, and they will be explored in a future work. For completeness, in Fig. \ref{fig:sim_density_radius} we show also the values obtained including the sink particles in the calculations. Figure \ref{fig:sim_mass_density} shows the integrated mass $M(R,\rho>\rho_{eff})$ within radius $R$, with filtering by a critical density. The blue line contains the total gas mass, and the orange, green, and red lines contain gas mass at density above $10^3$, $10^4$, $10^5$ cm$^{-3}$ , respectively, typical values of the diffuse gas observed in massive clumps. Solid lines show only the gas mass, and the dotted lines with same colours show the mass including sink particles. This figure shows that most of the gas is found at low densities and a lot of gas is missed during the observation if a critical density is introduced, probably biasing the measurement of the total gas kinetic energy of the system.

\section{Discussion}\label{app:discussion}
For a sample of virialized clumps, our starting hypothesis, the value of the \avir-mass slope is equal to zero by definition. 

As already noted by \citet{Kauffmann13}, the observed slope $h_{\alpha}$ depends on both the linewidth-size slope $\delta$ and the mass-radius slope $\gamma$ \citep[see in particular Eq. 10 in][]{Kauffmann13}:

\begin{equation}\label{eq:alpha_vir_mass_relation_exponent}
h_{\alpha}=\frac{2\delta+1-\gamma}{\gamma}.
\end{equation}

The artifacts in the observations therefore influence the estimation of the slopes $\gamma$ and $\delta$. While in reality the observational biases likely contribute to both these slopes, in the following we discuss the two extreme cases where they contribute to the evaluation of either $\gamma$ or $\delta$.

We first assume that the artifact affects the estimation of $\gamma$. In this case the bias is entirely due to the estimation of the gravitational energy of the system. The line width-size slope is well determined, the estimation of the kinetic energy no longer depends on $\rho_{eff}$ and our model, in particular Eq. \ref{eq:alpha_vir_mass_clump_relation}, does not apply. 

An observational bias in the estimation of the mass-radius slope may arise from the sensitivity limits intrinsic in each survey. Since within each survey the sensitivity remains the same, the mean column density of each clump remains very similar between different clumps. At the same time, the sensitivity can change significantly between different surveys, leading to different mean mass surface densities of a given sample \citep{Kauffmann13}.

A similar argument is discussed in \citet{Ballesteros-Paredes12}. For a given survey with sensitivity threshold near or right after the peak of the column density probability distribution of the clumps $N-$PDF, the average column density of the clumps is determined by the (relatively) low-density gas that dominates the distribution.

It is worth noting now that observations of massive clumps show that there is no correlation between line width and size; that is, the slope is $\delta\simeq0$ \citep{Ballesteros-Paredes11,Traficante17_PII,Traficante18}. Fixing the value $\delta=0$, the \a_eff-mass slope depends only on the mass-radius slope of a given sample. From Eq. \ref{eq:alpha_vir_mass_relation_exponent}, it follows that $h_{\alpha}\rightarrow 0$ when $\gamma\rightarrow 1$. As discussed in Section \ref{sec:data_results}, the values of $\gamma$ are systematically larger than 1. The observational bias suggested by \citet{Ballesteros-Paredes12} and \citet{Kauffmann13} in this case contributes to an over-estimation of the mass-radius slope.

On the other hand, if the observational bias is entirely due to the measurements of the velocity dispersion, the line width-size relation is the only one affected, implying $\delta\neq0$. In this case our model is fully responsible for the observed \a_eff-mass slope and $h=h_{\alpha}$. The bias in this hypothesis arises from the fact that the dust continuum surveys used to estimate the dust mass recover the dust emission of the whole structure, while any given gas tracer is only excited, in our simplified onion model, in layers at and above its effective critical density $\rho_{eff}$.

An example of how this bias affects real observation is the following: assuming a typical massive clump of 400 M\sun\ with average surface density of 0.1 g cm$^{-2}$ and radius $R=0.5$ pc, the average volume density is $\simeq1.2\times10^{4}$ cm$^{-3}$. From Eq. \ref{eq:rho_0_rho_average_relation} it follows that the volume density of the outer layer is $4\times10^{3}\leq\rho_{0}\leq8\times10^{3}$ cm$^{-3}$, assuming a radial density profile with $1\leq p\leq2$. In this example, any gas tracer with $\rho_{eff}\geq10^{4}$ cm$^{-3}$, the effective critical density of a commonly used high-density tracer, such as \n2h\ $(1-0)$ at T=10 K, will not be able to trace all the cold gas in this clump. We note that a gas tracer with a lower value of $\rho_{eff}$, such as the \nh3\ $(1,1)$ ($\rho_{eff}\leq10^{3}$ cm$^{-3}$ in gas at T=10 K), may suffer from the opposite bias: its emission can arise from all the low-density gas along the line of sight, including the gas in the parent filament/cloud \citep[e.g.][]{Battersby14} which cannot be disentangled from the clump-only emission.

Under the hypothesis that the bias is only due to the estimation of the velocity dispersion, the model can predict the correct power-law form expected for a sample of virialized clumps if the appropriate gas tracers are used to measure the velocity dispersion of clumps with different volume densities. Equation \ref{eq:alpha_vir_mass_relation_exponent} predicts that the correct value of $h_{\alpha}=0$ for a sample of virialized clumps would be obtained as a result of, for example, a line width-size relation with a slope $\delta=0.35$ and $\delta=0.69$ for the samples in \citet{Urquhart18} and \citet{Traficante18}, respectively, a result not far from what expected in the case of a turbulent cascade \citep{McKee07}.

\end{document}